\documentclass[a4paper]{article}

\usepackage{INTERSPEECH2022}
\usepackage{multirow}

\title{Fine-grained Noise Control for Multispeaker Speech Synthesis}
\name{Karolos Nikitaras$^1$, 
	  Georgios Vamvoukakis$^1$,
	  Nikolaos Ellinas$^1$,
	  Konstantinos Klapsas$^1$,
	  Konstantinos Markopoulos$^1$,
	  Spyros Raptis$^1$,
	  June Sig Sung$^2$
	  Gunu Jho$^2$,
	  Aimilios Chalamandaris$^1$,
	  Pirros~Tsiakoulis$^1$}
\address{$^1$Innoetics, Samsung Electronics, Greece\\
		 $^2$Mobile Communications Business, Samsung Electronics, Republic of Korea}
\email{k.nikitaras@partner.samsung.com, 
		\{g.vamvouk, n.ellinas\}@samsung.com,
		k.klapsas@partner.samsung.com,
		\{k.markop, s.raptis, js6.sung, gunu.jho, aimilios.ch, p.tsiakoulis\}@samsung.com}

\begin{document}
		\maketitle
	\begin{abstract}
		
		A text-to-speech (TTS) model typically factorizes speech attributes such as content, speaker and prosody into disentangled representations.
		Recent works aim to additionally model the acoustic conditions explicitly, in order to disentangle the primary speech factors, i.e. linguistic content, prosody and timbre from any residual factors, such as recording conditions and background noise.
		This paper proposes unsupervised, interpretable and fine-grained noise and prosody modeling.
		We incorporate adversarial training, representation bottleneck and utterance-to-frame modeling in order to learn frame-level noise representations.
		To the same end, we perform fine-grained prosody modeling via a Fully Hierarchical Variational AutoEncoder (FVAE) which additionally results in more expressive speech synthesis.
		Experimental results support our claims and ablation studies verify the importance of each proposed component.
		Audio samples are available in our demo page\footnote{https://innoetics.github.io/publications/fv-noise/index.html}.
		
	\end{abstract}


	\section{Introduction}
	
		Advances in text-to-speech (TTS) neural systems have resulted in models that are capable of synthesizing high quality speech from many different speakers.
		Such models can be autoregressive (AR) \cite{shen2018natural,shen2020non} or non-autoregressive (NAR) \cite{ren2020fastspeech, donahue2020end, kim2021conditional}, attentive \cite{shen2018natural} or duration-informed \cite{ren2020fastspeech, shen2020non, donahue2020end}, end-to-end \cite{kim2021conditional, donahue2020end} or not \cite{shen2018natural,shen2020non,ren2020fastspeech}.
		Non end-to-end systems depend on a vocoder \cite{oord2016wavenet,valin2019lpcnet,prenger2019waveglow} that synthesize speech from the predicted acoustic features.
		
		Multi-speaker TTS systems are commonly trained on publicly available corpora, such as VCTK \cite{yamagishi2019cstr} and LibriTTS \cite{zen2019libritts}.
		VCTK contains recordings from 109 speakers of different English dialects, whereas
		LibriTTS contains recordings from about 2,500 English speakers and is divided into disjoint subsets based on data quality.
		These corpora are collected in a large scale, having as a result multiple artifacts and noise.
		It is worth mentioning that one fourth of LibriTTS corresponds to signal-to-noise ratio (SNR) of 20 dBs or less \cite{zen2019libritts}, when approximated by waveform amplitude distribution analysis (WADA) \cite{kim2008robust}.
		Despite that fact, they are used widely in multiple tasks, such as voice cloning \cite{jia2018transfer,choi2020attentron} and prosody modeling \cite{sun2020fully}.
		
		In order to produce clean speech from such corpora it is important to explicitly model the noise \cite{hsu2019disentangling,lee2021styler,zhang2021denoispeech}.
		Furthermore, given that speaker identities and recording/noise conditions are often correlated \cite{hsu2019disentangling}, it is essential to learn speaker representations disentangled from acoustic conditions.
		Recent works aim to learn speech representations disentangled from noise-related factors \cite{hsu2019disentangling} or explicitly model the acoustic conditions \cite{hsu2019disentangling,lee2021styler,zhang2021denoispeech}.
		In both cases, speech and noise corpus mixing is a common practice \cite{hsu2019disentangling,lee2021styler,zhang2021denoispeech}.
		Typically, if the signal is mixed with artificial noise, it is considered as noisy, otherwise as clean.
		As a result, the inherent noise that potentially exists within signals is considered as part of the clean speech and thus, it cannot be explicitly modeled.
		Given that high-quality clean speech data is costly to collect, the only way to counteract such behavior is to incorporate unsupervised noise modeling.
		

	\subsection{Related Work}

	A simple way to improve the quality of a speech corpus is to use enhancement methods such as noise suppression and dereverberation prior to text-to-speech training \cite{8343873}.
	Recent works train the TTS model directly on the noisy data.
	Utterance-level speech and noise corpus mixing is a common technique and has been widely used.
	In \cite{hsu2019disentangling}, speaker timbre is disentangled from acoustic conditions and noise is modeled in a coarse grained way.
	
	Fine-grained noise modeling in a two-step way is performed in \cite{zhang2021denoispeech}, where a noise condition module based on U-Net \cite{ronneberger2015u} is trained jointly with the TTS model in a semi-supervised way.
	In Styler \cite{lee2021styler}, frame-level noise is considered as one of several style factors and is modeled in a fully supervised way.
	The latter two works both apply supervision over the noise representations as they require knowledge of whether the input signal contains clean or noisy speech.
	
	In terms of prosody, \cite{zhang2021denoispeech} models phoneme-level F0 and duration as it follows the FastSpeech architecture \cite{ren2020fastspeech}, while \cite{lee2021styler} also extends to phoneme-level energy modeling.
	Both methods use prosody labels extracted by external modules. Regarding the phoneme alignment, they work either in a supervised way or by simple linear compression or expansion of audio to the text's length respectively.
	We hope to extend previous work using a fully unsupervised training scheme both in terms of noise and prosody modeling.

	
	\begin{figure*}[!htbp]
		\centering
		\includegraphics[width=\textwidth]{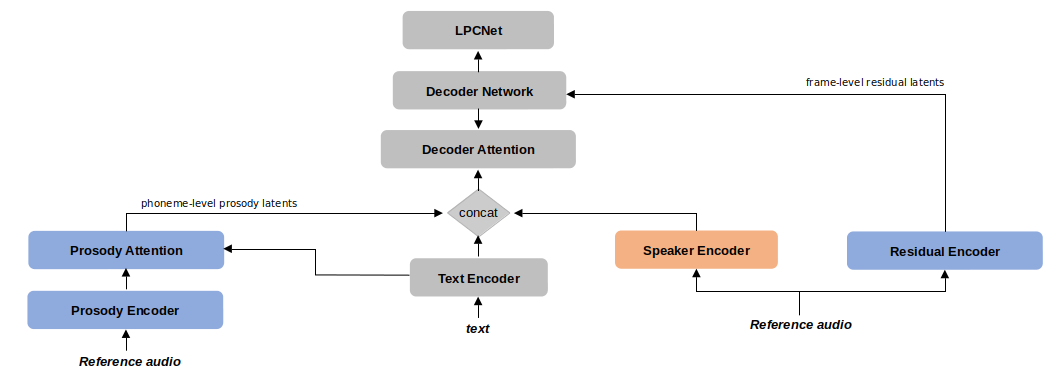}   
		\vspace*{-6mm}
		\caption{The overall architecture of the proposed system.}
		\label{fig:architecture}
		\vspace{-15pt}
	\end{figure*}

	\subsection{Proposed Method}
	
	We base our TTS model on a Tacotron-like architecture \cite{ellinas2021high}, which is an attentive, autoregressive seq-to-seq model and we closely follow \cite{hsu2019disentangling} in terms of the data augmentation process.
	In order to learn speech attributes other than content and speaker, we perform fine-grained noise and prosody modeling.
	Our conditional generative model can be written as $\boldsymbol{p}(speech\vert$$z_s,z_r,z_p,text)$, where $\boldsymbol{p}(z_r\vert$$speech)$ and $\boldsymbol{p}(z_p\vert$$speech,text)$ are the variational posterior distributions that correspond to the residual and prosody encoder, respectively, while $\boldsymbol{p}(z_s\vert$$speech)$ comes from pre-trained speaker encoder network (see section~\ref{section:speaker-encoder}).
	All the posterior distributions are defined as diagonal covariance Gaussian whose mean and variance are parameterized by each encoder network, while all the prior distributions as isotropic Gaussian.
	
	We propose a system that performs unsupervised, interpretable frame-level noise and phoneme-level prosody modeling. 
	In order to achieve disentanglement between speech and noise, we incorporate adversarial training, representation bottleneck and utterance-to-frame modeling. 
	Our prosody modeling also enhances this disentanglement, while results in more expressive speech synthesis.
	Experimental results show that unsupervised, fine-grained noise modeling outperforms other methods and this verifies the effectiveness of our system in controlling and so, removing even the inherent noise within the speech corpus.

	All in all, the contributions of this work are the following:
	
	
	\begin{itemize}
		\item unsupervised fine-grained noise modeling
		\item a novel utterance-to-frame training scheme that enables frame level noise modeling
		\item a low-dimensional (bottleneck) residual latent representation in addition to adversarial training in order to discourage linguistic content leakage into the residual latent space
		\item fine-grained prosody modeling to further enhance the disentanglement of the noise latent representations from speech related factors as well as synthesize expressive speech
		\item synthesizing speech with higher estimated SNR than the inherently noisy training data
		
	\end{itemize}

	The rest of the paper is organized as follows.
	In Section 2, we present the proposed system and its main components in terms of speaker \ref{section:speaker-encoder}, noise \ref{section:residual-encoder} and prosody \ref{section:prosody-encoder} modeling.
	Section \ref{section:experiments} provides ablation studies and objective evaluation comparing our proposed system to other alternatives regarding SNR estimation measured by WADA \ref{section:performance-evaluation} as well as the reconstruction performance \ref{section:reconstruction-performance}.
	Section \ref{section:conclusions} concludes the paper indicating directions for future work.

	\section{Method}	
	
	\subsection{Model Architecture}
	
	The overall model architecture is shown in Figure~\ref{fig:architecture}, which is based on \cite{ellinas2021high}.
	Text encoder converts the phoneme sequences into the text hidden sequence.
	Prosody encoder maps the mel-spectrogram sequence into frame-level hidden prosody sequence which in turn, is being aligned with the text hidden sequence.
	We then use the extracted alignment in order to produce the average, raw, phoneme-level mel representation.
	Fully Hierarchical Variational AutoEncoder (FVAE) takes as input the latter representation and produces hierarchically the three latent dimensions as in \cite{sun2020fully}.
	The projection of these three latent dimensions as well as the speaker embedding are concatenated to the text hidden sequence.
	The noise encoder operates in frame-level and produces the residual hidden sequence.
	Adversarial training is applied over the intermediate, frame-level residual representation.
	Also, the Variational Autoencoder (VAE) framework is being utilized over the latent representations of the prosody and residual encoder.
	Posterior distributions are defined as diagonal covariance Gaussian whose mean and variance are parameterized by each encoder network, while all the prior distributions as isotropic Gaussian with zero mean and unit variance.
	Finally, we condition the Decoder Network to the residual embedding at each step.
	
	\subsection{Speaker Encoder}
	\label{section:speaker-encoder}
	
	We use transfer learning from speaker verification as in \cite{jia2018transfer}.
	The pre-trained speaker encoder outputs fixed-dimensional embeddings called d-vectors, which are trained via the Generalized End2End (GE2E) loss \cite{wan2018generalized}.
	Trainable speaker encoder embeddings could have also been used\footnote{Some preliminary experiments showed that trainable speaker embeddings like \cite{hsu2019disentangling} are more robust to noise than pretrained d-vectors but are worse in terms of speaker similarity}, 
	however the focus of this work is noise modeling,
	hence we selected the pre-trained network for speaker embedding.

	\subsection{Residual Encoder}
	\label{section:residual-encoder}
	
	Unsupervised, frame-level modeling entails the risk of over-fitting. 
	Thus, we employ several techniques to counteract such behavior.
	The Residual Encoder has two blocks, each consisting of a convolutional and a batch normalization layer each.
	It maps the mel-spectrogram sequence into frame-level hidden residual representation.
	We impose a bottleneck over the residual representation by using a small number of dimensions.
	This prohibits the model to use the latent residual representation in order to directly pass frame-level spectral information to the decoder module.
	Additionally, adversarial training is being imposed over the hidden representation in order to further discourage the residual latent space from modeling linguistic content, by applying adversarial CTC loss \cite{graves2006connectionist}.
	We also perform utterance-to-frame modeling over the hidden sequence by applying mean pooling with a probability which starts from 1, at the early stage of training, and gradually decreases to 0.
	Otherwise, the system learns correlated speech and noise representation within the residual latent space, even when using just a single latent dimension.
	We further enhance this disentanglement by performing prosody modeling.
	Ablation studies in Section~\ref{section:performance-evaluation} verify the importance of each proposed component. 
	
	\subsection{Prosody Modeling}
	\label{section:prosody-encoder}
	
	We use three latent dimensions which are trained in a hierarchical, unsupervised way and correspond to phoneme-level f0, energy and duration as in \cite{sun2020fully}.
	Given that the noise existence (5-25 dB) within the augmented dataset is prohibitive to the phoneme-to-mel alignment extraction, we used the original version of the speech signals in the input of the prosody encoder.
	
	The Prosody Encoder passes the mel-spectrogram sequence through two blocks, each consisting of a convolutional and a batch normalization layer, followed by an LSTM layer.
	It outputs a hidden, frame-level prosody representation which is being aligned to the hidden text sequence via the Prosody Attention module.
	This alignment is being used along with the raw mel sequence resulting to an average, phoneme-level mel representation.
	FVAE takes in the latter representation and produces, in a hierarchical way, the posterior distributions of three phoneme-level, prosody latents.
	
	\section{Experiments}
	\label{section:experiments}
	
	Train-clean-100 is chosen as our train set which is a LibriTTS subset consisting of utterances with at least 20 dB SNR, measured by WADA \cite{zen2019libritts}. 
	The train set is augmented with background noise from the CHIME 4 challenge \cite{vincent2017analysis} following the data augmentation process in \cite{hsu2019disentangling}.
	The motivation here is not to counteract the inherent speaker-noise correlation in order to learn noise-invariant speaker representations as in \cite{hsu2019disentangling}, but to simulate real noisy data, impose noise-related attributes as the predominant residual information and be able to control and then remove the inherent noise existing in our data by not using noise labels in our fine-grained, noise modeling process.

	\subsection{Model and training setup}
	The input text is first normalized and converted into a phoneme sequence by a traditional TTS front-end module.
	The acoustic features used for training the synthesizer are matching the ones by the LPCNet vocoder \cite{valin2019lpcnet}, i.e. 20 bark-scale cepstral coefficients plus pitch period and pitch correlation, while sequences of 80-dim mel-scale filterbank frames are used as the input of the prosody and the residual encoders.
	
	The synthesizer network is based on the attentive, sequence-to-sequence architecture \cite{ellinas2021high}, with extra input $z_p$ as well as $z_s$ concatenated to the text encoder network output, while $z_r$ concatenated and passed to the decoder at each step.
	Assuming a batch sample of B utterances having P as the maximum number of phonemes and F as the maximum number of acoustic frames, then $z_s \varepsilon \mathbb{R}^{B x 256}$, $z_p \varepsilon \mathbb{R}^{B x P x 3}$ and $z_r \varepsilon \mathbb{R}^{Bx F x 1}$.
	
	The network parameters are trained with the Adam optimizer \cite{kingma2014adam}, which minimizes the average L1 loss before and after the post-net, batch size 32 and an initial learning rate of $ 3 * 10^{-3} $ linearly decaying to $ 3 * 10^{-5} $ after 100,000 iterations. L2 regularization with weight $ 10^{-5} $ is also used.
	
	\subsection{Residual Encoder Evaluation}
	
	The following three sections aim to quantify the ability of our system to model and control noise. First, we explore the mapping of several utterances with varying acoustic conditions into the residual latent space. We also estimate the SNR of the generated speech by traversing the frame-level latent space. Finally, we compare our system to several baselines in the clean speech production.
	
	\subsubsection{Training phase}
	
\begin{figure}[t]
		\centering
		\includegraphics[width=\linewidth]{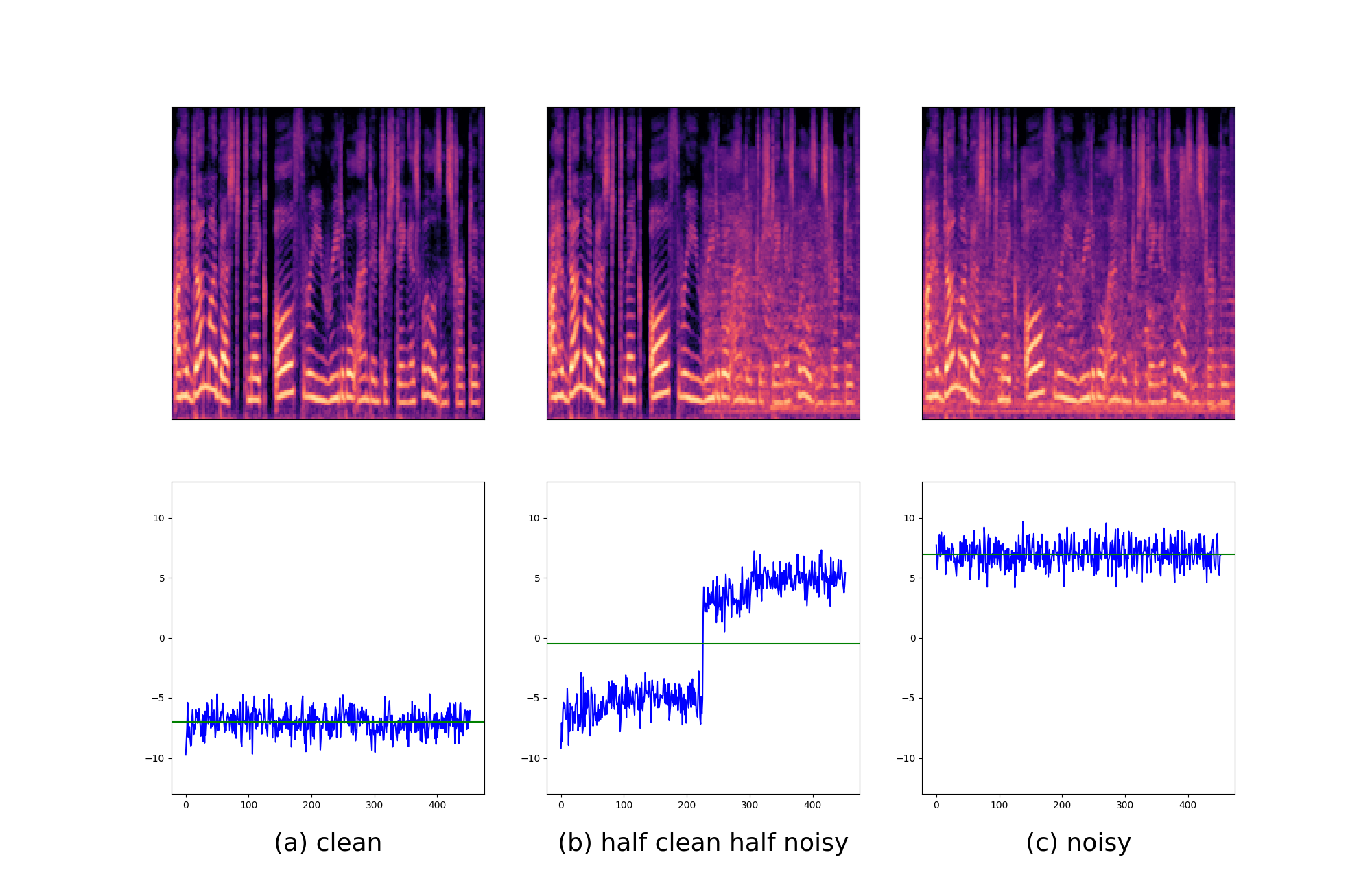}
		\caption{Utterances with varying acoustic conditions along with their residual latent representation}
		\label{fig:res-latent-space-mapping}
		\vspace{-15pt}
	\end{figure}

	Figure~\ref{fig:res-latent-space-mapping} illustrates a speech signal under different acoustic conditions, along with each residual latent representation. 
	Left column shows the clean version of the signal, at the center column artificial noise is added to the second half of the signal, while at the right column noise is added to the  entire length of the signal.
	It is shown that the 1-dimensional latent distribution consists of a clean and a noisy tail which correspond to the negative and the positive axis, respectively.
	
	\subsubsection{Inference phase}
	
\begin{figure}[t]
		\centering
		\includegraphics[width=\linewidth]{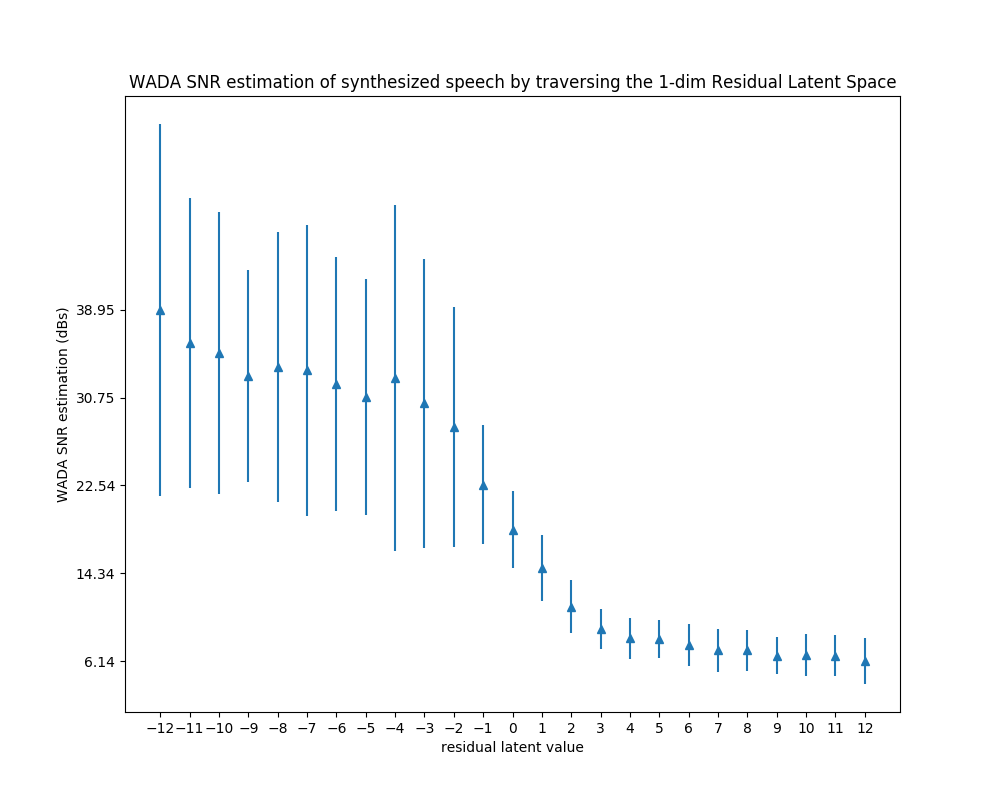}
		\caption{WADA-SNR estimation of synthesized speech by traversing the 1-dim Residual Latent Space}
		\label{fig:traversing-residual-latent-space}
	\end{figure}

	During inference time, our system needs to assign one residual value per frame that is being synthesized.
	Figure~\ref{fig:traversing-residual-latent-space} shows that by traversing the residual latent space in the range [-12,12], while broadcasting each value across all synthesizing frames, we generate speech and then estimate the mean SNR using WADA.
	The noisy tail corresponds to about 6 dBs which is close to 5 dB, the lowest SNR level that used in the data augmentation process.
	The high variance in the clean tail indicates that there is room for further improvements in order to achieve more robust noise modeling.

	\subsubsection{Performance Evaluation}
	\label{section:performance-evaluation}
	
	Results shown in Table~\ref{tab:snr-system} quantify the ability to control noise through WADA SNR estimation, by comparing the full proposed model with several alternative models, while also conducting ablation studies:
	\begin{enumerate}
		\item ``Ground-truth" corresponds to the ground truth recordings
		\item ``Ground-truth enhanced" are the ground truth recordings enhanced by a speech-enhancement model \cite{tzinis2020sudo}
		\item ``Noise-unaware" is a model trained on the ground-truth data
		\item ``Noise-unaware enhanced" is a model trained on the enhanced ground-truth data
		\item ``Utt-level" performs utterance-level noise modeling by closely following \cite{hsu2019disentangling}
		\item ``Proposed - CTC" is the proposed system without the adversarial CTC loss
		\item ``Proposed - utt-to-frame" is the proposed system without utterance-to-frame level modeling, that performs directly frame-level noise modeling
		\item ``Proposed - fvae" is the proposed system without the prosody modeling, trained on the artificially noisy corpus
		\item ``Proposed" is the proposed system.
	\end{enumerate}
	
	The systems (6,7,8,9) perform frame-level noise modeling,
	(5) performs utterance level noise modeling,
	while the rest do not perform any noise modeling.
	All systems are trained on the artificially noisy speech corpus, except from the two noise-unaware systems (4,5) that are trained on the ground-truth speech corpus.

	
	\begin{table}[]
		\caption{Average WADA-SNR of synthesized speech across several systems}
		\label{tab:snr-system}
		\centering

		\begin{tabular}{cc}
			\hline
			Model                   & mean WADA SNR 		   \\ \hline
			ground-truth            & 28.3                     \\
			ground-truth enhanced   & 28.7                     \\ \hline
			noise-unaware           & 21.5                     \\
			noise-unaware enhanced  & 21.1                     \\
			utt-level \cite{hsu2019disentangling}              & 21.3                     \\ \hline
			proposed - CTC          & 1.1                      \\
			proposed - utt-to-frame & 7.6                      \\
			proposed - fvae         & 23.2                     \\ \hline
			proposed                & \textbf{36.0}                     \\ \hline
		\end{tabular}
	\end{table}

	Without the adversarial CTC loss over the intermediate, frame-level residual latents, (6) fails to learn disentangled speech and noise representations.
	The same holds for the alternative that directly performs frame-level noise modeling (7).
	These two verify that unsupervised, frame-level modeling entails the risk of overfitting.
	Comparing the full proposed model (9) with the alternative without the FVAE (8), we show that by performing fine-grained prosody modeling we encourage the residual encoder to capture other factors than prosody and so, result in more clean speech.
	Supervised noise modeling (2,4) seems incapable of removing the inherent noise, since the performance between both (1,2) and (3,4) is quite close.
	Fine-grained noise modeling (8,9) results in more clean speech than coarse-grained (5).
	Finally, the difference between the proposed system (9) and the ground-truth recordings (1) illustrates the effectiveness of the unsupervised, fine-grained noise modeling in controlling and removing the inherent noise present in some of the ground-truth recordings.

	\subsection{Prosody encoder evaluation}
	\label{section:reconstruction-performance}
	
	Table~\ref{tab:prosody-performance} compares the reconstruction performance of the proposed system with several alternative models.
	To quantify the reconstruction performance we use mel-cepstral distortion (MCD) which evaluates the timbral distortion and F0 Frame Error (FFE) which evaluates the reconstruction of the F0 track.
	Lower is better for both metrics.
	The latter is a combination of the Gross Pitch Error (GPE) and the Voicing Decision Error (VDE).
	
	The ``Proposed - fvae" has slightly worse performance that ``utt-level" and this might be a result of the correlated speech and noise representation within the residual latent space.
	The Proposed model outperforms the two alternatives across all metrics.
	As in \cite{sun2020fully}, FVAE captures phoneme-level F0, energy and duration.
	The significant FFE drop indicates that F0 information is captured, while the drop in VDE that energy and duration are captured.
	
	\begin{table}[]
		\caption{Reconstruction performance results}
		\label{tab:prosody-performance}
		\centering
		\begin{tabular}{ccccc}
			\hline
			Model         & GPE           & VDE           & FFE           & MCD           \\ \hline
			utt-level     & 0.32          & 0.14          & 0.34          & 6.87          \\
			proposed-fvae & 0.38          & 0.14          & 0.36          & 6.96          \\
			proposed      & \textbf{0.30} & \textbf{0.11} & \textbf{0.28} & \textbf{6.03} \\ \hline
		\end{tabular}
	\end{table}

	\section{Conclusions}
	\label{section:conclusions}
	
	In this work, we propose a TTS system that achieves clean and expressive speech synthesis in an unsupervised and interpretable manner.
	We incorporate adversarial training, representation bottleneck and utterance-to-frame noise modeling in order to factorize noise and speech into disentangled representations.
	We also enhance this disentanglement by performing fine-grained prosody modeling, in the sense that we encourage the frame-level residual encoder to learn attributes other than prosody.
	Experimental results show that our system outperforms other methods in clean speech synthesis, indicating that our unsupervised, fine-grained noise modeling method can control and remove the inherent noise that exists in crowd-sourced data.
	In addition, for future work, we would also like to investigate the capability of the proposed method to learn richer noise representations in order to further improve the synthesized speech quality.
	
	\bibliographystyle{IEEEtran}
	
	\bibliography{mybib}
	
\end{document}